\documentclass[preprint,showpacs,preprintnumbers,amsmath,amssymb]{revtex4}

\usepackage{graphicx}
\usepackage{dcolumn}
\usepackage{bm}

\begin{document}

\title{One-Dimensional Three-State Quantum Walk}

\author{Norio Inui}
\email{inui@eng.u-hyogo.ac.jp}
\affiliation{%
Graduate School of Engineering, 
University of Hyogo, \\
2167, Shosha, Himeji, Hyogo, 671-2280, Japan\\}

\author{Norio Konno}
\email{norio@mathlab.sci.ynu.ac.jp}
\author{Etsuo Segawa}
\email{segawa820@lam.osu.sci.ynu.ac.jp}
\affiliation{%
Department of Applied Mathematics, 
Yokohama National University, 
79-5 Tokiwadai, Yokohama, 240-8501, Japan\\}

\date{\today}

\begin{abstract}
We study a generalized Hadamard walk in one dimension with three inner states. 
The particle governed by the three-state quantum walk moves, in superposition,  both to the left and to the right according to the inner state.
In addition to these two degrees of freedom, it is allowed to stay at the same position. We calculate rigorously the wavefunction of the particle starting from the origin for any initial qubit state, and show the spatial distribution of probability of finding the particle. In contrast with the Hadamard  walk with two inner states on  a line, 
the probability of  finding  the particle at the origin  does not converge to zero even after infinite time steps except special initial states. This implies that the particle is trapped near the origin after long time with high probability.
\end{abstract}

\maketitle

\section{\label{sec:S1} Introduction}

Quantum walks \cite{Aharonov1993,Mayer1996,Ambainis2001,Aharonov2001,Tregenna2002} are very simple quantum processes, however they are connected with a wide variety of field. Studies of the quantum walks are initially motivated by developing techniques for quantum algorithms. For example, the Grover's search algorithm \cite{Grover1997}, which is one of the most famous quantum algorithms, is related to a discrete quantum walk \cite{Shenvi2002,Ambainis2004}. Oka et al. \cite{Oka2005} studied  breakdown of an electric-field driven system by mapping the Landau-Zener transition dynamics to a quantum walk. 

The Hadamard walk plays a key role in studies of the quantum walk
 and it has been analyzed in detail. Thus the generalization of the Hadamard walk is one of fascinating challenges. The simplest classical random walker on a one-dimensional lattice moves to the left {\it or} to the right with probability $1/2$. On the other hand, the quantum  walker according to the Hadamard walk on a line  moves both to the left and to the right. It is well known that the spatial distribution of probability of finding a particle governed by  the Hadamard walk after long time is quite different from that of the classical random walk, see \cite{Kon2002,Kon2005,GJS,KFK}, for examples. However there are some commonalities. In both walks the probability of finding probability at a fixed lattice site converges to zero after infinite long time.

Let us consider a classical random walker in which it can stay at the same position with non-zero probability in a single time-evolution. 
In the limit of long time, dose the probability of finding a walker at a fixed position converges to a positive value ?  If the probability of staying at the same position in a single step is very close to 1,
the walker diffuses very slowly. However the probability of existence at a fixed position surely 
converges to zero if the jumping probability is not zero.
The classical random walker who can remain the same position is essentially regarded as the same process with the random walk 
right or left with probability with 1/2 by scaling the time.
Is the same conclusion valid for a quantum walk? The answer is ``No". 
We  show that the profile of the Hadamard walk changes
drastically by appending  only one degree of freedom
to the inner states.  

If the quantum particle in three-state quantum walk exists at only one site initially, the particle is trapped with high probability
near the initial position. Similar localization has been already seen in other quantum walks. The first simulation showing the localization was presented by Mackay et al. \cite{Mackay2002} in studying the two-dimensional Grover walk. After that, more refined simulations were performed by Tregenna et al. \cite{Tregenna2003} and an exact proof on the localization was given by Inui et al. \cite{Inui2004}. The second is found in the four-state quantum walk \cite{Brun2003B,Inui2005Physica}. In this walk, a particle moves not only the nearest sites but also the second nearest sites according to the four inner states. The four-state quantum walk is also a generalized Hadamard walk, and it is similar with the three-state quantum walk. The significant difference between them is that the wavefunction of the four-state quantum walk dose not converge, but that of  the three-state quantum walk converges in the limit as time tends to infinity. We focus on localized stationary distribution of the three-state quantum particle and calculate it rigorously. Moreover Konno \cite{Kon2005b} proved that a weak limit distribution of a continuous-time rescaled two-state quantum walk has a similar form to that of the discrete-time one. In fact, both limit density functions for two-state quantum walks have two peaks at the two end points of the supports. As a corollary, it is easily shown that a weak limit distribution of the corresponding continuous-time three-state quantum walk has the same shape as that of the two-state walk. So the localization does not occur in the continuous-time case in contrast with the discrete-time case. In this situation the main aim of this paper is to show that the localization occurs for a discrete-time three-state quantum walk rigorously. 

The rest of the paper is organized as follows. After defining the three-state quantum walk, eigenvalues and eigenvectors of the time evolution operator are calculated, and the wave function is shown in Sec. $\mbox{I\hspace{-0.5mm}I}$. A time-averaged probability of finding the particle is introduced in Sec. $\mbox{I\hspace{-0.5mm}I\hspace{-0.5mm}I}$, and  it is shown that the time-averaged probability of the three-state quantum walk converges to non-zero values. In Sec. $\mbox{I\hspace{-0.5mm}V}$, we prove the probability of finding the particle at a fixed itself converges to a non-zero value after infinite long time. 


\section{\label{sec:S2} Definition of the three-state quantum walk}

The three-state quantum walk (3QW) considered here is a kind of the generalized Hadamard walk on a line. The particle ruled by 3QW is characterized in the Hilbert space which is defined by a direct product of a chirality state space $|s \rangle \in \{ |L\rangle, |0\rangle, |R\rangle \}$ and a position space $|n \rangle \in \{\ldots, |-2 \rangle, |-1 \rangle, |0\rangle,|1 \rangle, |2 \rangle, \ldots \}$. The chirality states are transformed at each time step by the next unitary transformation:
\begin{eqnarray*}
|L\rangle = \frac{1}{3} (-|L\rangle+2|0\rangle+2|R\rangle ), \>\>
|0\rangle = \frac{1}{3} (2|L\rangle -|0\rangle+2|R\rangle ), \>\>
|R\rangle = \frac{1}{3} (2|L\rangle+2|0\rangle -|R\rangle ).
\end{eqnarray*}
Let $\Psi(n,t) \equiv [\psi_{L}(n,t),\psi_{0}(n,t),\psi_{R}(n,t)]^{T}$ be the amplitude of the wave function of the particle corresponding to the chiralities  ``L", ``0" and ``R" at the position $n \in \mathbb Z$ and the time $t \in \{0,1,2, \ldots \}$, where $T$ denotes the transpose operator and $\mathbb Z$ is the set of integers. We assume that a particle exists initially at the origin. Then the initial quantum states are determined by $[\psi_{L}(0,0),\psi_{0}(0,0),\psi_{R}(0,0)] \equiv [\alpha,\beta,\gamma],$ where $\alpha, \beta, \gamma \in \mathbb C$ with $|\alpha|^{2}+|\beta|^2+|\gamma|^{2}=1$. Here $\mathbb C$ is the set of complex numbers.

Before we  define the time-evolution of the wavefunction, we introduce the following three operators:
\begin{eqnarray*}
U_{L}=\frac{1}{3}\left [
\begin{array}{ccc}
-1 & 2 &  2 \\
0 & 0 &  0 \\
0 & 0 &  0 \\
\end{array}
\right ], \quad 
U_{0}=\frac{1}{3}\left [
\begin{array}{ccc}
0 & 0 &  0 \\
2 & -1 & 2 \\
0 & 0 &  0 \\
\end{array}
\right ], \quad
U_{R}=\frac{1}{3}\left [
\begin{array}{ccc}
0 & 0 &  0 \\
0 & 0 &  0 \\
2 & 2 &  -1 \\
\end{array}
\right ]. 
\label{eqn:U3}
\end{eqnarray*}
If the matrix $U_{L}$ is applied to the function $\Psi(n,t)$, the only ``L"-component is selected after carrying out the superimposition
between $\psi_{L}(n,t)$, $\psi_{0}(n,t)$ and $\psi_{R}(n,t)$. Similarly  the ``0"-componet and ``R"-componet 
are selected in $U_{0}\Psi(n,t)$ and $U_{R}\Psi(n,t)$.

We now define the time evolution of the wavefunction  by
\begin{eqnarray*}
\Psi(n,t) &=& U_{L} \psi(n+1,t)+U_{0} \psi(n,t)+U_{R} \psi(n-1,t).
\label{eqn:tev}
\end{eqnarray*}
One finds clearly that the chiralities  ``L" and ``R" are corresponding to the left and the right, 
and  the chirality ``0" is corresponding to the neutral state for the motion.


Using the Fourier analysis, which is often used in the calculations of quantum walks,
we obtain the wavefunction. The spatial Fourier transformation of $\Psi(k,t)$ is defined by
\begin{eqnarray*}
\tilde{\Psi}(k,t) =\sum_{n \in \mathbb Z} \Psi(n,t) e^{-i k n}.
\label{eqn:Fwf}
\end{eqnarray*}
The dynamics of wavefunction in the Fourier domain is given by
\begin{eqnarray}
\tilde{\Psi}(k,t+1) &=&  \frac{1}{3}\left [
\begin{array}{ccc}
e^{ik} & 0 &  0 \\
0 & 1 &  0 \\
0 & 0 & e^{-ik} \\
\end{array}
\right ]
\left [
\begin{array}{ccc}
-1 & 2 &  2 \\
2 & -1 &  2 \\
2 & 2 &  -1 \\
\end{array}
\right ]
 \tilde{\Psi}(k,t)  \nonumber \\
& \equiv &  \tilde{U} \tilde{\Psi}(k,t). 
\label{eqn:dyFou}
\end{eqnarray}
Thus the solution of Eq. (\ref{eqn:dyFou}) is formally given by $\tilde{\Psi}(k,t)=\tilde{U}^{t}\tilde{\Psi}(k,0)$.
Let $e^{i \theta_{j,k}}$ and $|\Phi_{k}^{j} \rangle$ be the eigenvalues of $\tilde{U}$ 
and the orthonormal eigenvector corresponding to $e^{i \theta_{j,k}}$ $(j=1,2,3)$. Since the matrix $\tilde{U}$ is a unitary matrix, it is diagonalizable. Therefore the wavefunction $\tilde{\Psi}(k,t)$ is expressed by
\begin{eqnarray*}
\tilde{\Psi}(k,t)=\left(\sum_{j=1}^{3} e^{i \theta_{j,k}t} |\Phi_{k}^{j} \rangle \langle\Phi_{k}^{j}  | \right)
\tilde{\Psi}(k,0),
\label{eqn:wf}
\end{eqnarray*}
where $\tilde{\Psi}(k,0) =[ \alpha, \beta, \gamma]^{T} \in \mathbb C^{3}$ with $ |\alpha|^{2}+|\beta|^2+|\gamma|^{2}=1.$ The eigenvalues are given by
\begin{eqnarray}
\theta_{j,k} &=& \left \{
\begin{array}{cc}
0, & j=1, \\
\theta_{k}, & j=2, \\
-\theta_{k}, & j=3, \\
\end{array}
\right . \nonumber \\
\cos \theta_{k} &=& -\frac{1}{3}(2+\cos k),  \nonumber \\
 \sin \theta_{k}&=& \frac{1}{3}\sqrt{(5+\cos k)(1-\cos k)}, 
\label{eqn:eigenvalue}
\end{eqnarray}
for $k \in [-\pi, \pi).$ The eigenvectors of $\tilde{U}$ which are orthonormal basis are obtained after some algebra:
\begin{eqnarray}
|\Phi_{k}^{j} \rangle &=& \sqrt{c_{k}(\theta_{j,k})}
\left [
\begin{array}{cc}
\frac{1}
     {1+e^{i(\theta_{j,k}-k)}}  
\\
\frac{1}
     {1+e^{i\theta_{j,k}}} 
\\
\frac{1}
     {1+e^{i(\theta_{j,k}+k )}}
\end{array}
\right ],
 \label{eqn:eigenV}
\end{eqnarray}
where 
\begin{eqnarray*}
c_k(\theta) 
= 2 
\left\{
\frac{1}{1+\cos(\theta-k)}+
\frac{1}{1+\cos \theta} + 
\frac{1}{1+\cos (\theta+k)} 
\right\}^{-1}.
\label{eqn:C}
\end{eqnarray*}
In the above calculation, it is most important to emphasize that there is 
the eigenvalue  ``1" independently on
the value of $k$. The eigenvalues of the Hadamard walk are given by $e^{i \theta_{k}}$ and
$e^{i (\pi-\theta_{k})}$  with the  arguments satisfying $\sin \theta_{k}=\sin k/\sqrt{2}$. Therefore the eigenvalues does not take the value ``1" except the special case $k=0$. We have shown the existence of strongly degenerated eigenvalue such as ``1" in Eq. (\ref{eqn:eigenvalue}) is necessary condition in the quantum walks showing the localization \cite{Inui2004}. The significant difference between 3QW and the previous quantum walks showing the localization such as the Grover walk and the four-state quantum walk is the number of the degenerate eigenvalues of the time evolution matrix. There are two degenerate eigenvalues with values ``1"  and  ``-1" independently of the value of $k$ in both the Grover walk and the four-state quantum walk. In contrast there is only one degenerate eigenvalue independently of the value $k$ for 3QW. This distinctive property causes the particular time evolution in 3QW.


Let us number each chirality ``L",``0", and ``R" using $l=1$, 2, and 3, respectively. Then 
the wavefunction in real space is obtained by the inverse Fourier transform: for $\alpha, \beta, \gamma \in \mathbb C$ with $|\alpha|^{2}+|\beta|^2+|\gamma|^{2}=1$, 
\begin{eqnarray}
\Psi(n,t; \alpha, \beta, \gamma)
&=& \frac{1}{2\pi} \int_{-\pi}^{\pi} 
\tilde{\Psi}(k,t) e^{ikn} dk \nonumber \\
&=& \frac{1}{2\pi} \int_{-\pi}^{\pi} 
\left( \sum_{j=1}^{3} e^{i \theta_{j,k}t} |\Phi_{k}^{j} \rangle \langle\Phi_{k}^{j}  |  \tilde{\Psi}(k,0) \right) e^{ikn} dk \nonumber \\
&=&
\left[\Psi(n,t;1;\alpha, \beta, \gamma), \Psi(n,t;2;\alpha, \beta, \gamma), \Psi(n,t;3;\alpha, \beta, \gamma) \right]^{T} \nonumber \\
&=& \sum_{j=1}^{3}
\left[\Psi_j (n,t;1;\alpha, \beta, \gamma), \Psi_j (n,t;2;\alpha, \beta, \gamma), \Psi_j (n,t;3;\alpha, \beta, \gamma) \right]^{T}, 
\label{eqn:realwfa}
\end{eqnarray}
where
\begin{eqnarray}
\Psi_j (n,t;l;\alpha, \beta, \gamma)
= \frac{1}{2\pi} \int_{-\pi}^{\pi} c_{k}(\theta_{j,k}) \varphi_{k}(\theta_{j,k},l) e^{i (\theta_{j,k}t+kn)} dk, \quad (l=1,2,3) 
\label{eqn:realwfb}
\end{eqnarray}
with
\begin{eqnarray}
&&
\varphi_{k}(\theta,l) 
= \zeta_{l,k}(\theta) 
[\alpha \overline{\zeta_{1,k}(\theta)} 
+ \beta \overline{\zeta_{2,k}(\theta)} 
+ \gamma \overline{\zeta_{3,k}(\theta)}], \quad (l=1,2,3) \nonumber \\
&&
\zeta_{1,k}(\theta) =(1+e^{i(\theta-k)})^{-1}, \quad
\zeta_{2,k}(\theta) =(1+e^{i \theta})^{-1}, \quad
\zeta_{3,k}(\theta) =(1+e^{i(\theta+k)})^{-1},
\label{eqn:realwfc}
\end{eqnarray}
where $\overline{z}$ is conjugate of $z \in \mathbb C.$ Sometimes we omit the initial qubit state $[\alpha, \beta, \gamma]$ such as $\Psi_j (n,t;l) = \Psi_j (n,t;l;\alpha, \beta, \gamma)$. The probability of finding the particle at the position $n$ and time $t$ with the chirality $l$ is given by $P(n,t;l)=|\Psi(n,t;l)|^{2}.$
Thus the probability of finding the particle at the position $n$ and time $t$ is 
$
P(n,t)=\sum_{l=1}^{3}P(n,t;l).
$


\section{\label{sec:S3} time-averaged probability}
We focus our attention on the spatial distribution of the probability of finding the particle after long time. Equation (\ref{eqn:realwfa}) is rather complicated, but the value of $\lim_{t \rightarrow \infty} P(n,t)$ can be exactly calculated in following sections. We start showing a numerical result for the probability of finding the particle at the origin before carrying  analytical calculation. Figure 1 shows the time dependence of $P(0,t)$ with the initial state $\alpha=i/\sqrt{2},\beta=0,\gamma=1/\sqrt{2}$. The probability decreases quickly near $t=0$ and fluctuates near 0.2. The inserted figure in Fig. 1 is plotted as the
inversed time, and the amplitude of fluctuation decreases as the time. The horizontal dashed line in Fig. 1 is the value at the origin of the time-averaged probability defined by
\begin{eqnarray*}
\bar{P}_{\infty} (0;\alpha,\beta,\gamma) &=& 
\lim_{N \rightarrow \infty} \left(\lim_{T \rightarrow \infty} \frac{1}{T}\sum_{l=1}^{3}\sum_{t=0}^{T-1} P_{N}(0,t;l;\alpha,\beta,\gamma) \right), 
\label{timeav}
\end{eqnarray*}
where $P_{N}(n,t;l;\alpha,\beta,\gamma)$ is the probability of finding a particle with the chirality $l$ at the position $n$ and time $t$ on a cyclic lattice containing $N$ sites. The time-averaged probability introduced here has already used to study the quantum walk. In the case of the Hadamard walk on a cycle containing odd sites, the time-averaged probability takes a value $1/N$ independently of the initial states \cite{Aharonov2001,Inui2005}. Therefore the time-averaged probability converges to zero in the limit $N \rightarrow \infty$. On the other hand the time-averaged probability of quantum walks which exhibits the localization converges to a non-zero value. For this reason we firstly calculate time-averaged probability of 3QW, and we will show in the next section that the probability $P(n,t)$ itself converges as well.

We calculate the time-averaged probability at the origin in 3QW on a cycle with $N$ sites. 
We assume that the number $N$ is odd. 
The argument of eigenvalues of 3QW with a finite $N$ is  $\theta_{j,2m\pi/N}$ for $m \in [-(N-1)/2,(N-1)/2]$.
Since  the eigenvalues corresponding to $m$ is the same with 
the eigenvalues corresponding to $-m$, the wavefunction is formally expressed by
\begin{eqnarray*}
\Psi_{N}(0,t;l;\alpha,\beta,\gamma) &=& \sum_{j=1}^{3}\sum_{m=0}^{(N-1)/2}c_{j,m,l}(N)e^{i \theta_{N,j,m} t},
\label{eqn:Product}
\end{eqnarray*}
where $\theta_{N,j,m}$ is $\theta_{j,2m\pi/N}$. We note here that the coefficients $c_{j,m,l}(N)$ depend on the initial state $[\alpha,\beta,\gamma]$, but we omit to describe it. Then the probability $P_{N}(0,t;\alpha,\beta,\gamma)$ at the origin is given by
\begin{eqnarray*}
P_{N}(0,t;\alpha,\beta,\gamma) &=& \sum_{l_{1},l_{2},j_{1},j_{2}=1}^{3} 
\sum_{m_{1},m_{2}=0}^{(N-1)/2}
c^{\ast}_{j_{1},m_{1},l_{1}}(N)
c_{j_{2},m_{2},l_{2}}(N)
e^{i (\theta_{N,j_{2},m_{2}}-\theta_{N,j_{1},m_{1}})t}.
\label{eqn:PNt}
\end{eqnarray*}
The coefficients $c_{j,m,l}(N)$ are determined from  the  product of eigenvectors. Although lengthy calculations are required to express the coefficients $c_{j,m,l}(N)$ explicitly, it is shown below that coefficients $c_{j,m,l}(N)$ except $j=1$ dose not contribute $P_{N}(t)$. Noting that the following equation
\begin{eqnarray*}
\lim_{T \rightarrow \infty} \frac{1}{T} \sum_{t=0}^{T-1} e^{i \theta t}
&=& 
\left\{
\begin{array}{cc}
1, & \theta=0, \\
0, & \theta \neq 0,
\end{array}
\right.
\end{eqnarray*}
we have
\begin{eqnarray}
\bar{P}_{N}(0;\alpha,\beta,\gamma) &=&  
\sum_{l=1}^{3} 
\left(
\left|\sum_{m=0}^{(N-1)/2}
c_{1,m,l}(N) \right|^{2} \right. \nonumber \\
&&\left.+
\left |\sum_{j=2}^{3} c_{j,0,l}(N) \right|^{2}+
\sum_{j=2}^{3} 
\sum_{m=1}^{(N-1)/2}
\left |c_{j,m,l}(N) \right|^{2}
\right).
\label{eqn:PNav}
\end{eqnarray}
The difference between the first term and the other terms is caused  by the difference of the degree of degenerate eigenvalues.

Let $\phi_{j,k}(N)$ be the eigenvectors corresponding to the eigenvalue $e^{i \theta_{j,k}t}$ of the time-evolution matrix of the 3QW with a matrix size $3N \times 3N$. They are easily obtained from $|\Phi_{k}^{j} \rangle$ by
\begin{eqnarray*}
|\phi^{j}_{m}(N)\rangle &=& \frac{1}{\sqrt{3N}} [ |\Phi_{m}^{j} \rangle,\omega|\Phi_{m}^{j} \rangle, \omega^{2}|\Phi_{m}^{j} \rangle,\ldots,
\omega^{N-1}|\Phi_{m}^{j} \rangle ],
\label{eqn:eigenV2}
\end{eqnarray*}
where $\omega=e^{2 \pi i/N}$. Since the coefficients $c_{j,m,l}(N)$ are proportion to the product of eigenvectors, the orders with respect to $N$ of the first term and the second term in Eq. (\ref{eqn:PNav}) are $\cal O$$(1)$ and $\cal O \mit( N^{-1})$, respectively. Thus we can neglect the second term in the limit of $N \rightarrow \infty$. Using the eigenvectors in Eq. (\ref{eqn:eigenV}), we have 
\begin{eqnarray*}
\bar{P}_{\infty} (0;\alpha,\beta,\gamma) &=&
(5-2 \sqrt{6})(1+|\alpha+\beta|^2+|\beta+\gamma|^2-2|\beta|^2).
\label{eqn:Pavorigin}
\end{eqnarray*}
The time-averaged probability takes the maximum value $2(5-2 \sqrt{6})$ at $\beta=0$, which is the value indicated by the horizontal dashed line in Fig. 1. The component of $\bar{P}_{\infty}(0;\alpha,\beta,\gamma)$ corresponding to $l=1,2,3$ are respectively given by  
\begin{eqnarray}
\bar{P}_{\infty}(0;1;\alpha,\beta,\gamma) &=& \frac{|\sqrt{6}\alpha-2(\sqrt{6}-3)\beta+(12-5\sqrt{6})\gamma|^{2}}{36}, \nonumber \\
\bar{P}_{\infty}(0;2;\alpha,\beta,\gamma)  &=& \frac{(\sqrt{6}-3)^{2}|\alpha+\beta+\gamma|^{2}}{9}, \nonumber \\
\bar{P}_{\infty}(0;3;\alpha,\beta,\gamma) &=& \frac{|\sqrt{6}\gamma-2(\sqrt{6}-3)\beta+(12-5\sqrt{6})\alpha|^{2}}{36}. 
\label{eqn:Pavorigin}
\end{eqnarray}
We stress here that the time-averaged probability is not always positive. If $\alpha=1/\sqrt{6}$, $\beta=-2/\sqrt{6}$ and $\gamma=1/\sqrt{6},$ then the time-averaged probability becomes zero, that is, $\bar{P}_{\infty}(0;1/\sqrt{6},-2/\sqrt{6},1/\sqrt{6}) = 0.$


\section{\label{sec:S4} Stationary distribution of the particle}

We showed that the time-averaged probability of 3QW converges to non-zero values except special initial states. This result, however, does not mean that a particle is observed with  almost the same probability at the origin after long time evolution. Indeed the probabilities of finding a particle in Grover walk and the four-state quantum walk, whose time-averaged probability converge to non-zero values, do not converge. Because the time-averaged probabilities in both cases depend on the parity of the time. Figure 1 suggests that the probability $P(n,t) = P(n,t; \alpha, \beta, \gamma)$ itself converges in a limit of $t \rightarrow \infty$. In this section we calculate the limit $P_{\ast}(n) \equiv \lim_{t \rightarrow \infty} P(n,t)$ rigorously and consider the dependence of $P_{\ast}(n)$ on the position $n$.

The wavefuntion given by Eq. (\ref{eqn:realwfa}) is infinite superimposition of the wavefunction $e^{i (\theta_{j,k}t+kn)}$. If the argument $\theta_{j,k}$ given in Eq. (\ref{eqn:eigenvalue}) for $j=2,3$ is not absolute zero, then the $e^{i \theta_{j,k}t}$ oscillates with high frequency in $k$-space for large $t$. On the other hand, $c_{k}(\theta_{j,k})\varphi_{k}(\theta_{j,k},l)$ in Eq. (\ref{eqn:realwfb}) changes smoothly with respect to $k$. Therefore we expect that the integration for $j=2,3$ in Eq. (\ref{eqn:realwfb}) becomes small as the time increases due to cancellation and converges to zero in the limit of $t \rightarrow \infty$. That is, 
\begin{eqnarray}
\lim_{t \to \infty} \sum_{j=2} ^3 \Psi_j(n,t;l;\alpha, \beta, \gamma) = 0,
\label{eqn:norio}
\end{eqnarray}
for $l=1,2,3$ and $\alpha, \beta, \gamma \in \mathbb C$ with $|\alpha|^{2}+|\beta|^2+|\gamma|^{2}=1$. This conjecture can be rigorously proved using the Riemann-Lebesgue lemma (see Appendix A). Consequently the probability $P_{\ast}(n)=P_{\ast}(n;\alpha,\beta,\gamma)$ is determined from the eigienvectors corresponding to the eigenvalue ``1", that is, $\theta_{1,k}=0$ (see Eq. (\ref{eqn:eigenvalue})), and  $l$-th component of $P_{\ast}(n;l)=P_{\ast}(n;l;\alpha,\beta,\gamma)$ is given by 
\begin{eqnarray}
P_{\ast}(n;l;\alpha,\beta,\gamma) = 
\left | \Psi_1 (n,t;l;\alpha, \beta, \gamma) \right |^{2}.
\label{eqn:Plimdef}
\end{eqnarray}
Note that $\Psi_1 (n,t;l;\alpha, \beta, \gamma)$ does not depend on time $t$, since $\theta_{1,k}=0$. By transforming integration in the left-hand side in Eq. (\ref{eqn:Plimdef}) into complex integral,
we have 
\begin{eqnarray}
P_{\ast}(n;1;\alpha, \beta, \gamma) &=& 
\left|
2 \alpha I(n) + \beta J_{+}(n) + 2\gamma K_{+}(n)
\right|^{2}, \nonumber \\
P_{\ast}(n;2;\alpha, \beta, \gamma) &=& 
\left|
\alpha J_{-}(n) + {\beta \over 2} L(n) + \gamma J_{+}(n) 
\right|^{2}, \nonumber \\
P_{\ast}(n;3;\alpha, \beta, \gamma) &=& 
\left|
2\alpha K_{-}(n) + \beta J_{-}(n) + 2 \gamma I(n) 
\right|^{2}, 
\label{eqn:Plim}
\end{eqnarray}
where $c=-5+2\sqrt{6} \> (\in (-1, 0))$ and 
\begin{eqnarray}
&&I(n) = \frac{2 c^{|n|+1}}{c^2-1}, 
\quad L(n) = I(n-1)+2I(n)+I(n+1), \nonumber \\
&&J_{+}(n) = I(n)+I(n+1), 
\quad J_{-}(n) = I(n-1)+I(n), \nonumber \\
&&K_{+}(n) = I(n+1), 
\quad K_{-}(n) = I(n-1), 
\label{eqn:integrals}
\end{eqnarray}
for any $n \in \mathbb Z.$ We should remark that it is confirmed $\bar{P}_{\infty}(0;l;\alpha,\beta,\gamma) = P_{\ast} (0;l;\alpha, \beta, \gamma)$ for $l=1,2,3$ by using Eqs. (\ref{eqn:Pavorigin}), (\ref{eqn:Plim}) and (\ref{eqn:integrals}). 

Here we give an example. From  Eqs. (\ref{eqn:Plim}) and (\ref{eqn:integrals}), we obtain 
\begin{eqnarray}
&&
P_{\ast}(0;i/\sqrt{2},0,1/\sqrt{2})=
{4 c^2 (5 c^2 + 2c + 5) \over (1-c^2)^2}= 10 - 4 \sqrt{6} = 0.202 \ldots, 
\\
\label{eqn:yuki}
&&
P_{\ast}(|n|;i/\sqrt{2},0,1/\sqrt{2})=
{2 (5 c^4 + 2 c^3 + 10 c^2 + 2c + 5) \over (1-c^2)^2} c^{2|n|},
\label{eqn:aki}
\end{eqnarray}
for any $n \ge 1$. Furthermore, 
\begin{eqnarray}
0 < \sum_{n \in  \mathbb Z} P_{\ast}(n;i/\sqrt{2},0,1/\sqrt{2}) =1/\sqrt{6} 
= 0.408 \ldots < 1. 
\label{eqn:kazue}
\end{eqnarray}
That is, $P_{\ast}(n;i/\sqrt{2},0,1/\sqrt{2})$ is not a probability measure. The above value depends on the initial qubit state, for example,
\begin{eqnarray*}
&& \sum_{n \in  \mathbb Z} P_{\ast}(n;1/\sqrt{3},1/\sqrt{3},1/\sqrt{3}) = 3 - \sqrt{6} = 0.550 \ldots, \\ 
&& \sum_{n \in  \mathbb Z} P_{\ast}(n;1/\sqrt{3},-1/\sqrt{3},1/\sqrt{3}) = (3 - \sqrt{6})/9 = 0.061 \ldots.
\end{eqnarray*}
We should remark that in the case of the classical symmetric random walk starting from the origin, it is known that $P_{\ast}(n) = 0$ for any $n \in \mathbb Z,$ therefore we have $\sum_{n \in  \mathbb Z} P_{\ast}(n)=0.$ The same conclusion can be obtained for the discrete-time and continuous-time two-state quantum walks \cite{Kon2002,Kon2005,Kon2005b}.

Figure 3 shows the probability $P_{\ast}(n;i/\sqrt{2},0,1/\sqrt{2})$ in log scale. The probability decreases exponentially for large $|n|$ and its asymptotic behavior is express by $P_{\ast}(n;i/\sqrt{2},0,1/\sqrt{2}) \propto c^{2|n|}$ in the limit of $n \rightarrow \pm \infty$ as Eq. (\ref{eqn:aki}) indicates.

We here mention the time-dependence of the $P(n,t)$. The particle is observed near the origin with high probability. This, however, dose not imply that the particle can not escape form the region near the origin. Figure 4 shows the change of $P(n,t)$ in a space-time. The particle is observed with high-probability  at the dark region. One clearly finds three dark regions. The first is a region near a center line connecting with the origin, and we can confirm the localization
near the origin. The second and third regions are boundaries of the triangle.
These regions show trajectories of two peaks moving outside in the space-time,  which are also seen in the Hadamard walk. As a result  we conclude that the particle in 3QW  splits three parts in superposition.

\section{\label{sec:S5} Conclusions and discussions}

The unique properties which are not observed in other quantum walks were found in three-state quantum walk. The particle which exists at the origin splits three pieces in superposition. Two of three  leave for infinite points and the remainder stays at the origin. In contrast with the ordinary  Hadamard walk, the probability of finding the particle at the origin does not vanish for large time, and its maximum probability is $10 - 4 \sqrt{6} = 0.202 \ldots.$

A simple reason why the 3QW is different form other quantum walks is the difference in the degree of degenerate eigenvalues. The necessary condition of the localization is the existence of the degenerate eigenvalues. And furthermore, the each degree of degeneration must be proportion to the dimension of the Hilbert space. In addition, if the degenerate eigienvalue is 1 only, then the probability of finding the particle can converges in the limit of $t \rightarrow \infty$. The Grover walk and the four-state quantum walk exhibit the localization, but the probability of finding the particle oscillates. Because there are degenerate eigenvalues with values ``1" and ``-1". On the other hand, the degenerated eigenvalue in 3QW unrelated to the wave number is only ``1", therefore the probability converges. Although no experiment exists about the quantum walks yet, the stationary properties of 3QW may be advantage to comparison with theoretical results.

Finally we discuss a relation between the limit distribution for the original 3QW $X_t$ as time $t \to \infty$ and that of the rescaled $X_t/t$ in the same limit. When we consider the 3QW starting from a mixture of three pure states $[1,0,0]^T, [0,1,0]^T,$ and $[0,0,1]^T$ with probability $1/3$ respectively, we can obtain a weak limit probability distribution $f(x)$ for the rescaled 3QW $X_t/t$ as $t \to \infty$ in the following:
\begin{eqnarray}
f(x) = {1 \over 3} \delta_0 (x) + { \sqrt{8} \> I_{(-1/\sqrt{3},1/\sqrt{3})}(x) \over 3 \pi (1-x^2) \sqrt{1-3 x^2}},
\label{eqn:pocky}
\end{eqnarray}
for $x \in [-1,1]$, where $\delta_0 (x)$ denotes the pointmass at the origin and $I_{(a,b)}(x)=1,$ if $x \in (a,b), \> =0,$ otherwise. The above derivation is due to the method by Grimmett et al. \cite{GJS}. We should note that the first term in the right-hand side of Eq. (\ref{eqn:pocky}) corresponds to the localization for the original 3QW. In fact, from Eq. (\ref{eqn:integrals}), we have  
\begin{eqnarray*}
{1 \over 3} \sum_{n \in  \mathbb Z} 
\left[ P_{\ast}(n;1,0,0) + P_{\ast}(n;0,1,0) + P_{\ast}(n;0,0,1) \right] = {1 \over 3}.
\end{eqnarray*}
The last value $1/3$ is nothing but the coefficient $1/3$ of $\delta_0 (x).$ Moreover, the second term in the right-hand side of Eq. (\ref{eqn:pocky}) has a similar form a weak limit density function for the same rescaled Hadamard walk with two inner states starting from a uniform random mixture of two pure states $[1,0]^T$ and $[0,1]^T$: 
\begin{eqnarray*}
f_{H}(x) = { I_{(-1/\sqrt{2},1/\sqrt{2})}(x) \over \pi (1-x^2) \sqrt{1-2 x^2}},
\label{eqn:pockyko}
\end{eqnarray*}
for $x \in [-1,1]$, see \cite{Kon2002,Kon2005,GJS}. This case does not have a delta measure term corresponding to a localization. More detailed study on this line will appear in our forthcoming paper.

\appendix
\section{Proof of Eq. (\ref{eqn:norio})}
Here we give an outline of the proof of Eq. (\ref{eqn:norio}). A direct computation gives 
\begin{eqnarray*}
\sum_{j=2} ^3  
\left[
\begin{array}{c}
\Psi_j(n,t;1;\alpha, \beta, \gamma)  \\
\Psi_j(n,t;2;\alpha, \beta, \gamma)  \\
\Psi_j(n,t;3;\alpha, \beta, \gamma)  \\
\end{array}
\right] 
=M 
\left[
\begin{array}{c}
\alpha \\
\beta  \\
\gamma \\
\end{array}
\right],
\end{eqnarray*}
where $M=(m_{ij})_{1 \le i,j \le 3}$ with
\begin{eqnarray*}
&& m_{11} = 3 J_{n,t} +{1 \over 2} 
\left\{ J_{n-1,t} + J_{n+1,t} + \left( K_{n-1,t} - K_{n+1,t} \right) \right\},
\\
&& m_{33} = 3 J_{n,t} +{1 \over 2} 
\left\{ J_{n-1,t} + J_{n+1,t} - \left( K_{n-1,t} - K_{n+1,t} \right) \right\},
\\
&& m_{12} = - 
\left\{ J_{n,t} + J_{n+1,t} + \left( K_{n,t} - K_{n+1,t} \right) \right\},
\\
&& m_{32} = - 
\left\{ J_{n,t} + J_{n-1,t} + \left( K_{n,t} - K_{n-1,t} \right) \right\},
\\
&& m_{13} = - 2 J_{n+1,t}, \qquad m_{31} = - 2 J_{n-1,t},
\\
&& m_{21} = - 
\left\{ J_{n,t} + J_{n-1,t} + \left( K_{n-1,t} - K_{n,t} \right) \right\},
\\
&& m_{23} = - 
\left\{ J_{n,t} + J_{n+1,t} + \left( K_{n+1,t} - K_{n,t} \right) \right\},
\\
&& m_{22} = 4 J_{n,t},
\end{eqnarray*}
and
\begin{eqnarray*}
J_{n,t} &=& \frac{1}{2\pi} \int_{-\pi}^{\pi} 
\frac{\cos(kn)}{5+\cos k} \cos (\theta_k t) dk,
\\
K_{n,t} &=& \frac{1}{2\pi} \int_{-\pi}^{\pi} 
\frac{\cos(kn)}{\sqrt{(5+\cos k)(1-\cos k)}} \sin (\theta_k t) dk.
\end{eqnarray*}
 From the Riemann-Lebesgue lemma, we can show that 
\begin{eqnarray*}
\lim_{t \to \infty} J_{n,t} =0, \qquad 
\lim_{t \to \infty} (K_{n,t} - K_{n+1,t})=0,
\end{eqnarray*}
for any $n \in \mathbb Z$. Therefore we have the desired conclusion.

\clearpage

\begin{figure}
\includegraphics{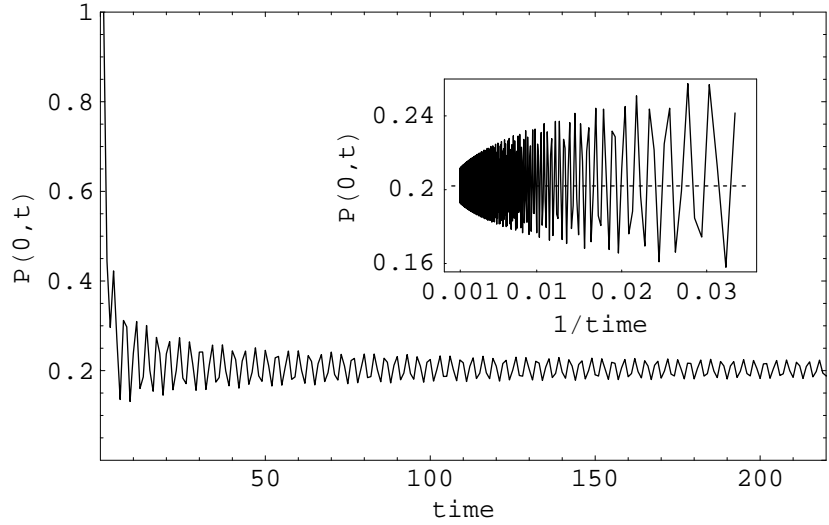}
\caption{\label{fig:Figure1} 
Time dependence of the probability of finding a particle at the origin starting 
from an initial state $\alpha=i/\sqrt{2},\beta=0,\gamma=1/\sqrt{2}$.
The same data are plotted as a function of  $1/t$ in the inserted figure.
The horizontal dashed line cross at  $(0,2(5-2 \sqrt{6}))$. The amplitude of fluctuation of probability decreases
near zero.
}
\end{figure}

\clearpage

\begin{figure}
\includegraphics{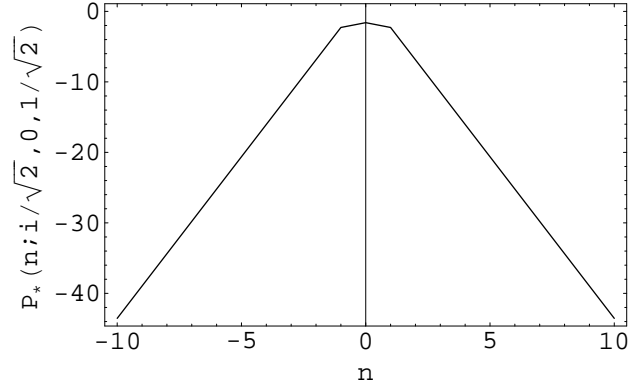}
\caption{\label{fig:Figure2} 
Distribution of probability of finding a particle $P_{\ast}(n)$ at the position $n$ in 3QW with the initial state $\alpha=i/\sqrt{2}, \beta=0$ and 
$\gamma=1/\sqrt{2}$ on a semi-log scale. The probability decreases exponentially for large $|n|$.
}
\end{figure}

\clearpage

\begin{figure}
\includegraphics{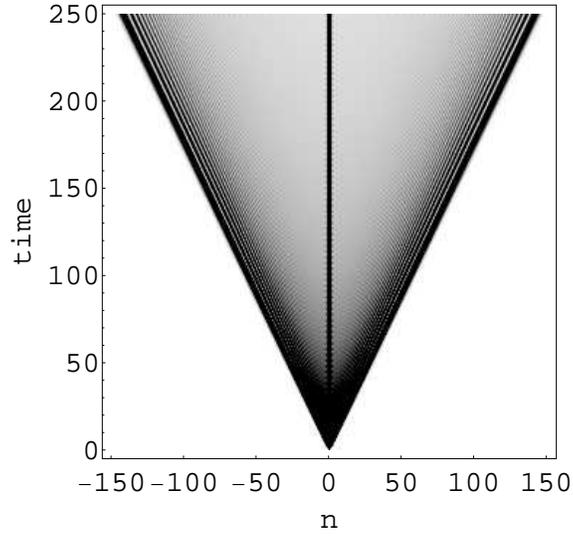}
\caption{\label{fig:Figure3} 
Density plot of probability of finding a particle $P(n,t)$ in the space-time.
The initial state is  $\alpha=i/\sqrt{2},\beta=0,\gamma=1/\sqrt{2}$.
The particle is observed on darker regions with high probability. 
}
\end{figure}

\clearpage

\end{document}